\documentclass[12pt,preprint]{aastex}

\begin{document}

\title{Star Formation History of a Young Super-Star Cluster in NGC 4038/39: Direct Detection of Low Mass Pre-Main Sequence Stars}

\author{Julia Greissl}
\affil{Steward Observatory, University of Arizona, Tucson, AZ 85721}
\email{jgreissl@as.arizona.edu}

\author{Michael R. Meyer}
\affil{Steward Observatory, University of Arizona, Tucson, AZ 85721}

\author{Micol H. Christopher}
\affil{California Institute of Technology, Pasadena, CA 91125}

\author{Nick Z. Scoville}
\affil{California Institute of Technology, Pasadena, CA 91125}

\begin{abstract}
We present an analysis of the near-infrared spectrum of a young massive star cluster in the overlap
region of the interacting galaxies NGC 4038/39 using population synthesis models. Our goal is to model the
cluster population as well as provide rough constraints on its initial mass function (IMF). The cluster
shows signs of youth such as thermal radio emission and strong hydrogen emission lines in the near-infrared.
Late-type absorption lines are also present which are indicative of late-type stars in the cluster.
The strength and ratio of these absorption lines cannot be reproduced through either late-type pre-main sequence (PMS) stars
or red supergiants alone.
Thus we interpret the spectrum as a superposition of two star clusters of different ages, which is feasible since the
1" spectrum encompasses a physical region of $\approx$ 90 pc and radii of super-star clusters are generally 
measured to be a few parsecs. One cluster is young ($\leq$ 3 Myr) and is responsible
for part of the late-type absorption features, which are due to PMS stars in the cluster,
 and the hydrogen emission lines. The second cluster is older (6 Myr - 18 Myr) and is needed
to reproduce the overall depth of the late-type absorption features in the spectrum. Both are required 
to accurately reproduce the near-infrared spectrum of the object. 
Thus we have directly detected PMS objects in an unresolved super-star cluster for the first time
using a combination of population synthesis models and pre-main sequence tracks.
This analysis serves as a testbed of our technique to constrain the low-mass IMF in young
super-star clusters as well as an exploration of the star formation history of young UC HII regions.
\end{abstract}

\keywords{galaxies: individual (NGC 4038, NGC 4039) -- galaxies: interactions -- stars: mass function -- stars: pre-main sequence}

\section{Introduction}

The advent of the Hubble Space Telescope (HST) has made it possible to study in detail nearby starburst galaxies with
star-formation rates orders of magnitude higher than the Milky Way
(e.g. M82, Henize 2-10, NGC 5253) and many are now known to host massive (10$^5$ - 10$^6$ M$_{\odot}$) young super-star clusters (SSC).
These clusters are thought to mirror a mode of star-formation that was ubiquitous in the early universe (\citet{leit01})
as evidenced by the fact that star formation rates of galaxies at high redshift are very similar to those in local starbursts \citep{stei96}.

SSC represent an extreme form of star formation which cannot be studied in the Milky Way. The closest analogue is
R136 in the LMC with a mass of 10$^{4.5}$ M$_{\odot}$ \citep{mahu98} though the Galaxy also hosts some clusters with masses greater than 10$^4$ M$_{\odot}$ (e.g. Westerlund 1, NGC 3603). Massive young clusters have been of particular interest in the search for IMF variations.
Since the local stellar IMF (within 1 kpc) appears to be universal \citep{mey00} we are forced to expand the search for IMF variations to more extreme regions of
star formation. SSC often form in the intense radiation environment of nuclear starbursts and interacting galaxies.
Thus they represent ideal objects to search for IMF variations with initial conditions such as metallicity and formation environment. Many young massive clusters with ages $\leq$ 40 Myr
have varying mass-to-light ratios (e.g. NGC 1705-1 \citep{smg01}, NGC 1569-B \citep{and04}),
which have often been interpreted as changes in the mass function. The mass-to-light ratios are determined through
dynamical mass estimates however, which assume that clusters are in virial equilibrium.
\citet{goba06} have shown that this assumption can be faulty in young clusters due to the effects of gas expulsion.
The mass-to-light ratio of M82-F however, a massive cluster in the prototypical starburst M82 with an age of 40-60 Myr, remains inconsistent with a \citet{kro01}
IMF \citep{mcc05}. Mass-to-light ratios are at best an indirect way to determine the IMF of a young cluster because the whole stellar mass range
is represented by one mass bin.
A more direct method of detecting stellar populations in unresolved SSC is needed to provide confirmation of the shape of their IMF.

Star clusters are dominated dynamically by their low-mass content. Yet, most of the light in young clusters is seen
through their massive stars. To understand whether a young SSC will remain bound or disperse it is important to get a
direct census of low-mass stars. In addition differences in the IMF of SSCs are most easily detected in low-mass stars, since the characteristic mass in both a \citet{kro01} and \citet{cha03} IMF is around 0.5 M$_{\odot}$.
Even small variations in the IMF of a young SSC may make the difference between a bound and an unbound cluster. The difference
between a Salpeter (1955) and a Chabrier (2003) IMF accounts for a factor of two in the mass of low-mass stars, for example.



The Antennae (NGC 4038/9) are the nearest (distance = 19.2 Mpc, 1" $\approx$ 93 pc, \citet{whi99}) pair of merging spiral galaxies.
They contain up to a thousand young star clusters. These clusters have been
the target of a multitude of studies at different wavebands (e.g. \citet{whi99}; \citet{neff00}; \citet{bran05}). \citet{whi95}
 identified a large population of SSC in the optical using HST WFPC2 images. The faintest and reddest of
these sources are revealed to be bright near-infrared and mid-infrared emitters and contain
the youngest clusters (\citet{snij06}; \citet{wan04}; \citet{bran05}). Radio images by \citet{neff00},
which detect sources with purely thermal nebular emission (i.e. $\alpha_{4cm-6cm}$ $\geq$ -0.4) due to massive stars in the clusters, provide additional evidence for the youth of these objects. Sources with non-thermal radio emission likely contain supernova remnants and are thus older.

This paper expands on a method introduced in Meyer \& Greissl (2005) to constrain the low-mass stellar
content in young SSC using late-type absorption features in the near-infrared. For young clusters most low-mass
objects are still on the pre-main sequence (PMS) and are thus orders of magnitude brighter than their main sequence counterparts. Thus they can be detected in
high signal-to-noise spectra of young SSC. We focus our analysis on one massive young cluster in the Antennae.

Cluster 89/90 (designation of \citet{whi95}), the cluster targeted in this study, was first surveyed by \citet{whi95} using WFPC2
on the HST. It is the brightest near-infrared cluster in NGC 4038/9 as well as the second brightest thermal
radio source at 4 and 6 cm. In addition \citet{snij06} observed the cluster in the mid-infrared and found strong NeII 12.8 and NeIII 15.5
$\micron$ emission as well as PAH emission at 11.25 $\micron$. These features are an indication of the presence of hot stars \citep{mir98}.
See Table 1 for a list of broadband magnitudes for cluster 89/90 obtained from archival data.

In the following sections we present a K- and H-band spectrum of cluster 89/90 and
model its underlying population. Section 2 details the observations and data reduction. Section 3 gives an overview of the method used
to model the spectrum of the cluster.
The analysis of the spectrum, which constrains the population of the cluster, is presented in section 4. Section 5
contains the results of the analysis, as well as its limitations. Section 6 places our analysis in the context of previous
work. Section 7 details our conclusions.

\section{Observations and Data Reduction}

H- and K-band spectra of cluster 89/90 were obtained with NIRSPEC \citep{mcl98} on Keck in February 2003 using the
42" slit with a scale of 0.144"/pixel. The spectrum was observed as part of a larger dataset of Antennae clusters
which are described in detail in \citet{chr08}. For the H-band spectra we used the N5 filter, covering 1.54-1.83 $\micron$.
The K-band spectrum was obtained with the N7 filter covering 2.05-2.47 $\micron$.
The K-band seeing during the observations was $\approx$ 0.5" and stable throughout the night; accordingly the 0.57" wide slit was used in both bands. The spectra
were obtained with a total integration time of 900 s (3 x 300s) in both the H- and the K-band at an airmass of $\approx$ 1.3.
The object was offset along the slit in successive frames to allow for sky subtraction.
Since SSC can often have a wealth of small-scale structure we were careful to make sure
the sky subtracted images contained no significant galactic background contamination.
After sky subtraction the background had no identifiable shape and the noise in the background was dominated by instrumental noise.
We did not see detectable CO absorption outside of our cluster in the spectrum which would have affected our analysis.
Calibration data were obtained including a flatfield and appropriate dark frame as well as neon and argon arc lamp spectra to use for wavelength calibration.
To correct for telluric absorption we used a G2V star in the H-band and an A0V star in the K-band. To ensure the best telluric correction we observed the standard
stars within 0.05 airmasses of the target.
Flatfielding and cosmic ray removal were performed on each spectrum using standard IDL procedures.


NIRSPEC spectra have spatial and spectral distortions that must be removed during the reduction process.
The spatial distortion corrections were calculated by measuring the position of the brightest calibrator spectra at multiple positions in the slit.
The traces of these sources were fit with polynomials to determine the spatial distortion correction. This routine was modified
from the REDSPEC\footnotemark[1] package written by Lisa Prato. For wavelength calibration, we used an argon arc lamp in the H-band and a neon
arc lamp in the K-band. \\
\footnotetext[1]{See http://www2.keck.hawaii.edu/inst/nirspec/redspec/index.html}
The reduction of the calibrator stars was carried out in the same manner.
To determine the atmospheric calibration we applied template spectra using \citet{mey98} in the H-band and \citet{wal97} in the K-band
 smoothed to the resolution of our spectra (R $\approx$ 1100 - 1500).
The calibrator stars were largely featureless, except for Br$\gamma$ absorption in the K-band,
which was not well matched by the template and removed independently.  Stellar absorption centers were
measured in both the template and the calibrator spectrum to remove any spectral offset.
The H-band template spectra covered the entire wavelength range of the H-band atmospheric calibrator observations; for the K-band, the template spectra ended at 2.4 $\micron$
while our observations extended redward to 2.47 $\micron$. We assumed a featureless blackbody for the template spectra from 2.4-2.47 $\micron$.
We examined the final telluric spectrum to ensure there were no mismatches between the G2V standard and the solar spectrum
particularly around the Mg absorption line at 1.71 $\micron$ which could have affected our analysis.
The atmospheric calibration was calculated by dividing the calibrator spectrum by the shifted template spectrum and normalizing the result at the center of the band.
Any spectral offset between the atmospheric correction and the source spectra were measured by comparing the location of atmospheric absorption features and applying an offset to the atmospheric correction if necessary. Typical offsets were less than 0.1 pixels.
The atmospheric correction removes both the effects of the atmosphere and variations in NIRSPEC system throughput as a function of wavelength.
The final spectrum was extracted from the atmospherically calibrated composite exposure with an aperture size of 5 x the seeing. Figure 1 shows the final extracted H- and K-band spectrum of cluster 89/90.

\section{Modeling the Near-Infrared Spectrum of Unresolved Star Clusters}

We model the spectra of young super-star clusters in the near-infrared incorporating the fact that
at young ages low-mass stars are still on the PMS. For objects on the PMS we use our own synthesis code combined with a set
of PMS tracks \citep{sie00}. For any objects not on the PMS we use the STARBURST99 (S99) population synthesis
models \citep{leit99} which are designed to accurately reproduce spectrophotometric properties of starbursts and SSC but do not incorporate any PMS tracks. For our S99 models
we used the included Padova tracks with AGB stars and the atmospheres used were Pauldrach/Hillier \citep{vale05}.
The final model has two separate components, one stellar component as well as a nebular component due to thermal free-free and free-bound emission from hot stars in the cluster.
In section 3.1 we describe the inputs and parameters of the stellar model, while 3.2 details the constraints we can place
on the nebular emission in clusters empirically.

\subsection{Stellar Component}

The stellar spectrum of our clusters is modeled in the following way (see also \citet{megr05}).
To determine the most massive star still on the PMS at a given age, we use the tracks of \citet{sie00}.
At ages of 1 and 3 Myr this corresponds to 7 M$_{\odot}$ and 5 M$_{\odot}$ respectively. An appropriate PMS
mass-luminosity relationship for this age is then assumed \citep{sie00}. Stars above this limiting mass are modeled
through S99. This includes the main sequence as well as the post-main sequence. We use instantaneous burst models assuming a \citet{sal55}
IMF above the PMS cutoff combined with the evolutionary models of the Padova group.
The mass of the PMS and S99 components
are scaled according to the IMF used and the total mass of our simulated clusters is 10$^6$ M$_{\odot}$. We adopt
a lower mass cutoff of 0.08 M$_{\odot}$ and an upper mass cutoff of 100 M$_{\odot}$. We have adopted solar metallicity
for our analysis in the Antennae in accordance with metallicity measurements (\citet{men01}; \citet{chr08}).

Assuming an IMF and an age, all stars below the PMS cutoff in the simulated cluster are assigned a standard star spectrum (\citet{mey98}; \citet{wal97})
depending on their temperature, with an appropriate luminosity. The standards used are field dwarfs which cover a spectral range between B2 and M6. The spectra
are scaled with the appropriate luminosity, coadded and weighted by the main sequence and post-main sequence contribution from S99.

\subsection{Nebular emission}

In addition to stellar light the near-infrared spectrum also contains nebular free-free and free-bound emission due to ionizing radiation emitted by the massive stars present in high-mass clusters. S99 predicts that at very young ages up to 90 $\%$ of the near-infrared luminosity
of young SSC should be emitted in nebular mission. However, the free-free and free-bound components can be constrained directly through
thermal radio data. Cluster 89/90 was observed by \citet{neff00} at 4 and 6 cm (Table 1). Using the 4 cm flux, which is less susceptible to non-thermal
contamination we can estimate the Lyman continuum flux being emitted by the cluster in the H and K band in
the following way. The number of Lyman continuum photons is related to the radio flux by the following
formula \citep{con92}: \\
\ \\
\begin{eqnarray}
\label{Qlyc} Q_{\rm Lyc} \geq 6.3\times10^{52}
\left({T_e \over 10^4{\rm ~K}}\right)^{-0.45}\times \left({\nu \over {\rm GHz}}\right)^{0.1}\\
\times \left({F_{nebular} \over 10^{27} {\rm ~erg ~s^{-1} ~Hz^{-1}}}\right)
\nonumber
\end{eqnarray}
\\

where T$_e$ is the electron temperature.
We assumed a value of  7500 K for T$_e$ which is a 'typical' temperature assumed for Galactic UC HII regions and
the same temperature assumed by \citet{neff00}. For cluster 89/90 this results in Q$_{Lyc}$ $\geq$ 8.46 x $10^{52}$ $s^{-1}$ which corresponds to
the equivalent of $\geq$ 2300 O5 stars powering the free-free and free-bound
emission of the cluster. We then convert this to a nebular flux in the H- and K-band following:

\begin{eqnarray}
\label{Ltherm}	{F_{nebular}} = \left({c \over \lambda^2}\right) \times \left({\gamma_{total} \over {\alpha_B}}\right) \times Q_{Lyc}
\end{eqnarray}
\\

Here, $\alpha_B$ refers to the case B recombination coefficient and $\gamma_{total}$ is the continuous emission
coefficient. The values used were adopted from \citet{fer80}. With the assumption of a distance for the
Antennae this value can then be directly compared to the measured near-infrared flux for cluster 89/90 in the H-band and the K-band (see Table 1). This results in
F$_{nebH}$ =  2.78 x $10^{-13}$ and F$_{nebK}$ = 1.98 x $10^{-13}$ erg s$^{-1}$ $\micron^{-1}$ $cm^{-2}$ in cluster 89/90. This corresponds to
 19.1 $\%$ and 23.5 $\%$ of the total H- and K-band flux from the cluster. This is the value we will use for
the rest of the analysis as the nebular percentage contributing to the near-infrared emission in cluster
89/90. The slope of the free-free and free-bound emission in the near-infrared is f$_{ff}$ $\propto$ $\lambda^{-2}$. S99 assumes that
every Lyman continuum photon emitted in the cluster is absorbed and reradiated into free-free or free-bound emission (case B). This assumption seems to overestimate the
measured nebular emission in young SSC. We attribute this to the fact that the dust surrounding the hot stars in young clusters is
likely clumpy. \citet{ind06} have found that the near-infrared SEDs of high-mass stars can vary by
orders of magnitude depending on the clumpiness of the circumstellar material.

\section{Application to Cluster 89/90 in the Antennae}

We now model the spectrum of cluster 89/90 comparing the NIRSPEC spectrum to the results of the modeling routine described above. Figure 1
shows the H- and K-band spectrum of the cluster with emission lines marked. The absorption lines are barely visible at this scale. Emission
lines are not included in our model or our analysis.
To model the spectrum of cluster 89/90 we first assume that the cluster represents one co-eval burst of star formation (see 4.1) and explore
the IMF required to reproduce the spectrum. We also consider two separate bursts as the underlying population of the spectrum (see 4.2), since
i) the spatial scale covered by our spectrum is quite large with $\approx$ 50 $\%$ of the clusters in the
 sample of \citet{chr08} revealed as multiple clusters in HST imaging and ii) the IMF required by a single co-eval burst exceeds the Salpeter slope.
Table 2 lists the best-fit model parameters for the two approaches.

\subsection{Single-age burst}

To constrain the IMF required in cluster 89/90 if the late-type absorption features are due to a single-age burst we model clusters at ages of 0.3, 1,3  and 5 Myr.
Above 6 Myr red supergiants appear in clusters and our single-burst model would no longer be appropriate due to the observed ratios of near-infrared absorption lines in our spectrum (as explained in more detail in the section 4.2). We found that the 5 Myr models
could not accurate reproduce
the observed spectrum because of the rapid dimming of the PMS stars and these models are not included in the two-burst model for that reason.
We use a single variable power-law for the IMF with a break-point at 1 M$_{\odot}$ with a variable slope below the break point and a \citet{sal55} slope ($\alpha$ = 2.35 in
linear units) above. We vary
the slope below 1 M$_{\odot}$ by 0.5 dex between 0.0 to 4.0. To determine which models best fit the cluster spectrum we performed a $\chi^2$ analysis comparing the model spectrum convolved to the resolution of the data and the data spectrum, which has been shifted to restframe wavelengths and normalized by fitting a 5th-order polynomial to the continuum.
The nebular emission in the cluster is calculated as given above from the radio data of \citet{neff00} and is independent of these assumptions.
We included five spectral regions in our fit, which are marked in Figure 2. These five regions include late-type absorption lines that are seen in
red supergiants as well as PMS stars. In the H-band there are numerous $^{12}$CO transitions which are surface
gravity sensitive and strong in red supergiants but weak in main sequence  and PMS objects. Of these we include the transitions at 1.62 $\micron$ and 1.71 $\micron$ in our fit.
The CO(2-0) 2.29 $\micron$ transition in the K-band is also surface gravity sensitive but
is a prominent feature in all late-type objects. The Ca triplet (2.261, 2.263 \& 2.117 $\micron$) and the Na doublet (2.206 \& 2.208 $\micron$) in the K-band are strongest in stars below 3500 K ($\leq$ 0.5 M$_{\odot}$). In addition we included the MgI 1.71 $\micron$ feature which is blended with one of the $^{12}$CO features at 1.71 $\micron$ and is strongest in stars between 0.5 - 1.5 M$_{\odot}$ (3500 - 7000 K). The presence of the Mg (solar-type stars) line in combination with the Ca, Na and CO (cooler stars) features is what allows us to place constraints on the IMF \citep{megr05}.
We then minimize $\chi^2$ of the difference between the model spectrum and the data. Here the goodness of the fit is given by: $\chi^2$ = $\sum_{i=0}^n (d_i - m_i)/\sigma^2_i$, where d$_i$ refers to the ith data pixel and m$_i$ refers to the corresponding model point and $\sigma^2_i$ is the signal-to-noise value at the ith pixel. $\chi^2$ is reduced
in the standard way by diving by the number of pixels in our absorption bands minus the degrees of freedom in our analysis.
This process is described in more detail in the following section.

The best-fit model is plotted in Fig. 2 and the model parameters are listed in Table 2. This single-age model does not match the depth of the surface gravity
sensitive CO features well and the spectrum which best matches the data has dN/dM $\propto$ m$^{-3.0}$ or steeper.
As this seems extraordinary, we consider an alternate hypothesis of two separate bursts of star formation at different ages.

\subsection{Two separate bursts}

We now attempt to reproduce the spectrum given two underlying bursts of star formation.
Given that our 1" slit represents a physical scale of $\approx$ 90 pc at the distance of the Antennae this approach seems warranted.
We model the spectrum as one young population containing PMS stars (hereafter Population 'A') in addition to an older population containing
red supergiants (hereafter Population 'B'). We model these populations separately and then combine the two. Incorporating an older cluster containing supergiants
is supported by the fact that the
H-band spectrum of cluster 89/90 contains $^{12}$CO absorption lines which are much stronger in supergiants
than in dwarfs and PMS stars (see Fig. 2 \& 4).
However, Figure 3 shows that cluster 89/90 lies below supergiant standards when using an index that is surface gravity sensitive as function
of temperature. Giants lie between supergiants and dwarfs
in this index but are not included in the figure since giants do not appear until ages $\geq$ 100 Myr.
This index includes the surface gravity sensitive CO(2-0) 2.29 $\micron$ feature as well as the NaI doublet and CaI triplet which are not strongly affected by surface gravity, though the ratio of the lines in the NaI doublet changes with log(g).
For reference, supergiant sources have log(g)= 0.0 while dwarfs have log(g)= 5.0 - 5.5. Cluster 89/90 lies close
to the dwarf locus in this index which indicates that the late-type absorption features in cluster 89/90
cannot be caused solely by red supergiants, but must have an underlying PMS component. \footnotemark[2]
\\\footnotetext[2]{The values
measured for Cluster 89/90 for the equivalent widths in Fig. 3 are: CO(2-0) = 14.84 $\pm$ 3.29 $\AA$, Ca = 3.30 $\pm$ 0.48 $\AA$ and
Na = 4.19 $\pm$ 0.63 $\AA$. The Ca triplet and Na doublet bandpasses
were chosen in accordance with \citet{klha86} while the CO(2-0) bandpass was 2.293 - 2.318 $\micron$ with the continuum between 2.2887-2.2915 $\micron$.}
The presence of a young component is
also supported by the emission lines seen in the spectrum of cluster 89/90.
The red supergiants have stronger late-type absorption lines but cannot match the observed surface gravity sensitive equivalent width ratios.
A combination of red supergiants and late-type dwarfs however can reproduce both the surface gravity index as well as the overall depth of the absorption features.
In our best-fit model A and B have a flux ratio of $\approx$ 7 to 1 which is well reproduced by the location of Cluster 89/90 in Figure 3.

There are three variables we consider for each burst: mass, age and IMF. The nebular emission in the cluster is again estimated from the
radio data of \citet{neff00}. The spectrum of B is modeled using S99, with appropriate standard spectra (\citet{mey98}; \citet{wal97})
used for the red supergiants in the spectrum similar to A. The output of S99 includes the number of stars of each
spectral type. For each supergiant of a given spectral type and temperature we then use an appropriate spectral standard.
The temperature scale used was adapted from the cool supergiant temperature scale by \citet{lev05}.

\subsubsection{Model Parameters}

According to S99 the first supergiants in a star cluster appear around 6 Myr though a significant population does not appear before 7 Myr. Late-type supergiants
disappear at an age of $\approx$ 30 Myr. Thus we can roughly constrain the age of B to between 7 and 30 Myr. In the H-band, cool stars possess a $^{12}$CO absorption line which blends with the 1.71 $\micron$ Mg line in the H-band (see Figure 2). This feature is strongly surface gravity sensitive and very weak in dwarfs. This blend is observed in Cluster 89/90 giving additional credence to the fact that red supergiant features are present in our spectrum. We considered models in 3 Myr steps.

For A the goal was not to constrain the age accurately but to assure that the best fit was consistent across different ages. Changing results with age would have limited the usefulness of our analysis method, since we do not have an independent constraint
on the age of A except for an upper limit on the age of $\leq$ 5 Myr.
We considered models of 0.3, 1 and 3 Myr.

We vary the mass ratio of the two populations between zero and one in steps of 0.01 below M$_{old}$/M$_{young}$ = 0.2 and in steps of 0.1 above 0.2. We cannot directly constrain the total mass of the cluster, because the overall flux of the spectral model
depends on the mass ratio of the best-fit model. After we have determined the best spectral model fit it is possible to constrain the mass through broadband magnitudes.
There is no way to directly constrain the IMF of B since we only trace a very small mass range of stars in the population in our spectrum, the red
supergiants. Thus the IMF of B is assumed to be the same as that of A.
For A we have allowed the IMF to vary as a broken power-law with a Salpeter slope above 1 M$_{\odot}$ and a varying slope below by 0.5 dex between 0.0 and 2.0 and also including 2.35. We did not attempt varying the break point in the IMF or introduce a low-mass cutoff. There is some
evidence that one could expect the Jeans mass to vary in SSCs which would influence the break point of the IMF \citep{lar05}.
The limitations of this particular data set, due to a possible overlap of two star clusters, make such a study not feasible.

\subsubsection{Analysis}
We again perform a $\chi^2$ analysis between the model spectra and the data including the same spectral regions as in 4.1.
To assess the signal-to-noise ratio (SNR) of the observed spectrum at each spectral region we fit the continuum of the spectrum
around each of the absorption bands and measured the RMS noise around the fit. We estimate the average SNR of the spectrum to be $\approx$ 30.
We ran models in all permutations of parameters described in the previous section and determined the best-fit model (with the lowest $\chi^2$) which is
shown in Figure 4.

Given the large grid of models, we tried to constrain the acceptable range of values for each model parameter in the following way. First we determined the probability of each model given by P$_i$ $\propto$ $e^{-{\chi_i}^2}$ and normalized so that $\sum_{i=0}^n$ P$_i$ = 1. We then plot the probability separately for each variable (see Figure 5),
where each point in the plot represents the sum of the probabilities over the whole range of the other two variables. We also assessed the probabilities in a Monte Carlo way. We added a noise spectrum given by a resampled subtraction of the data and the best-fit model to the cluster spectrum and determined the best fit for each newly realized resampled spectrum. We performed this routine 10000 times and determined the probability a particular set of parameters is determined as the best-fit model. The results from the Monte Carlo simulations agreed well with the probabilities determined from the $\chi^2$ values alone.

\section{Results}

\subsection{Best-fit Models}

We now discuss the implications of the ranges of parameters constrained by our modeling (see Table 2). The three plots in Figure 5 show the probability distribution
of the age of the old population, mass ratio of the two bursts as well as the slope of the IMF below 1 M$_{\odot}$. Each plot shows the probability distribution at the three different ages of A to
make sure that the results are consistent over a range of ages.
Panel 1 shows that the older population has a most probable age of 12 Myr. Within 90$\%$ confidence limits the models are consistent with an age between 6 - 18 Myr for the old population.
It should be noted that a population containing no supergiants is ruled out as can be seen by the fact that an age of 1 Myr and 3 Myr for B have zero probability.
This coarse age constraint is not surprising since very little differentiates spectra of red supergiants in the near-infrared besides the overall strength of late-type absorption lines.

The plot of mass ratios shows the cumulative probability with respect to M$_{old/young}$. The mass ratio strongly favors A dominating in mass by a factor of $\geq$ five
over B. The best-fit model comparison of the K-band flux with the K-band broadband flux of the cluster yields a total mass of 1.4 x $10^7$ M$_{\odot}$ down to the hydrogen burning limit. If we assume that the total mass is 1.4 x $10^7$ M$_{\odot}$ then the total mass of A is $\geq$ 1.2 x $10^7$ M$_{\odot}$, while the mass for B is $\leq$ 2 x $10^6$ M$_{\odot}$. The best-fit indicates a mass of B of 5 $\times$ 10$^5$ M$_{\odot}$ and a ratio of M$_{old}$/M$_{young}$ = 0.04. The total mass estimate agrees well with that calculated by \citet{gil00}
 (1.6 x $10^7$ M$_{\odot}$) using the Lyman continuum flux of cluster 89/90. Thus the young component of cluster 89/90 likely is one of the most massive young star clusters which formed in the Antennae galaxies.
The 90 $\%$ confidence limit for the mass ratio is roughly 0.02 - 0.12 though this value changes slightly between different ages of A. Thus we have clearly detected the PMS in the cluster. A mass ratio of 0 and therefore M$_{old}$ = 0 is ruled out as well as a mass ratio $\geq$ 0.12 and thus no PMS contribution.

The probability for the IMF of A rises strongly towards steeper slopes with the probability being highest for a Salpeter slope below 1 M$_{\odot}$. However the spectrum is
formally consistent with a slope down to a power-law slope of 1.5 within a 90$\%$ confidence limit. This result is consistent across all ages of the young population. A top-heavy IMF weighted more heavily towards high-mass stars than a \citet{kro01} IMF has often been cited as expected for SSC clusters and the best remaining evidence of an unusual IMF in a SSC (M82-F) indicates a top-heavy IMF. Our result is consistent with a normal Galactic IMF \citep{cov08} as well as a Salpeter IMF. A low-mass cutoff in this
cluster is ruled out since PMS objects in the young burst below 1.0 M$_{\odot}$ are required to produce the observed spectrum.
We did not include any slopes above Salpeter as the goal of this study was not
to make claims of extraordinary IMF results within the limited data this cluster offers. In addition, the best-fit model for the two-burst model at a Salpeter slope
shown in Figure 4 is a very good fit to the data whereas for the single burst even a Salpeter slope did not fit the data well.
This is independent of the probability curve shown which was shown to illustrate the overall trends of the models rather than the quality of the overall
fits. Thus we feel it is reasonable to set the cutoff at the Salpeter slope even though the IMF probability is still rising.

\subsection{Caveats}

Our model has many components and therefore degeneracies exist between its different subsets. One such degeneracy exists between the assumed age of the PMS objects and the IMF slope. Since PMS
objects grow fainter as they age one might erroneously assign an IMF that is too flat to a 1 Myr cluster that is really 3 Myr old. However, Figure 5
shows that our IMF results are very similar at different assumed ages for the young burst and thus we do not expect this to be a problem.
Another concern is that the calculated nebular emission is wrong and this might
affect the results of our models. We varied the input nebular continuum by 50$\%$ in both directions which did not change our results substantially.

In addition there are limits to the inputs of our models. Our standards are
field dwarfs with high surface gravity, while PMS objects are young and have lower surface gravity than dwarfs. The surface gravity of PMS objects is generally between log(g) = 3.0 - 4.2 \citep{gor03}, while dwarfs are generally have log(g) = 5.0 - 5.5. This would increase the depth of the CO absorption features in our spectrum. Obtaining spectral
standards of young PMS objects is difficult because the largest nearby sample is found in relatively distant young clusters, making it observationally expensive to obtain a complete high signal-to-noise sample in the near-infrared. Since we used spectral standards to model our spectra rather than synthetic spectra our coverage of individual
spectral types is not complete. This might cause errors in the depth of the absorption features of our model spectra due to binning the stars into spectral types for which spectral coverage exists. However, this effect is likely small compared to the error introduced by the usage of dwarf rather than PMS standards. Since synthetic spectra struggle to accurately model low-mass young PMS stars \citep{djw03}, currently dwarf standards remain the best option to accurately model the near-infrared spectra of SSC.
However, PMS stars generally lie closer to field dwarfs than supergiants in surface gravity \citep{gor03} and it is thus likely that even including PMS standards would not
remove the need of having a supergiant component in our models.
The presence of veiling due to disks around young stars can dilute the absorption features in near-infrared spectra. We do not have an independent method of quantifying the effect veiling has on our data since we are modeling a whole star cluster but we expect the effect to be minimized by the use of line ratios. Finally in some very young objects, CO features can be seen in emission due to the presence of a disk (e.g. \citet{blum08}). We do not observe evidence for this in our spectrum.

\section{Discussion}

\subsection{Previous Work}

We have shown that it is possible to directly detect PMS stars in unresolved SSC as well as place some constraints on the underlying IMF and age of the cluster.
Substantial work has been done studying the IMF of SSC through measurements of mass-to-light ratios, including in the Antennae. \citet{men02} measured mass-to-light ratios in the
K-band for a number of clusters in NGC 4038/9 and found variations indicative of IMF variations in the clusters. To measure mass-to-light ratios in clusters younger than $\approx$
30 Myr requires the ability to constrain velocity dispersions through strong CO absorption bands that exist in clusters dominated by red supergiants.
\citet{bas06} used UV spectra of clusters in two merger remnants with ages of more than 300 Myr to measure their mass-to-light ratios and found them to be consistent with a Kroupa IMF. Similarly \citet{lar04} and \citet{lari04} measured mass-to-light ratios in older clusters through optical spectroscopy and found them to be consistent with a Kroupa IMF. This
dichotomy: young clusters with varying mass-to-light ratios and older cluster with mass-to-light ratios consistent with normal IMFs provides evidence that something other than IMF
variations are the cause of varying mass-to-light ratios in young SSC or that perhaps SSC with unusual IMFs get disrupted preferentially. This work shows an avenue for providing more concrete measurements of IMFs in SSC than mass-to-light
measurements can deliver though this dataset is not the ideal application of our method.

Work on modeling the integrated spectra of young clusters has up until now been done mostly in the UV and optical (e.g. \citet{tre01}) because of the strong signatures of massive stars at these wavelengths. The near-infrared is ideally suited to detect the low-mass stellar content of the clusters as well as evolved giants and supergiants. Recently
\citet{lan08} have attempted to use simple stellar population models to constrain ages of star clusters in M82 in the near-infrared. \citet{bran05} attempted to use near-infrared colors of clusters in the Antennae in combination with thermal radio images to constrain their properties and found them to be uncorrelated. However, Antennae clusters are not ideal objects to study young clusters in detail. They are distant enough that multiple stellar populations of different ages may be treated as unresolved point sources. We have shown that it is possible to constrain the properties of young star clusters through near-infrared spectroscopy. Varying nebular emission due to the ionizing radiation emitted by hot stars provides an additional complication in the understanding of the youngest clusters. The total amount of free-free and free-bound emission in young clusters is easiest to constrain empirically, because it should depend strongly on the formation environment of the cluster. More work is needed
studying nearby SSC in the near-infrared to understand whether near-infrared images and spectroscopy can accurately predict the ages and masses of these clusters.

\subsection{Scales of Star Formation}

What can cluster 89/90 tell us about the scales of star formation in the Antennae galaxies as well as locally? The spectrum of our cluster likely contains two separate underlying
populations with an age spread between 6-18 Myr and a mass ratio M$_{old}$/M$_{young}$ $\leq$ 0.2. At the distance of the Antennae our 1" slit covers a region of 93 pc. This region likely contains
two massive clusters, one with a mass of order 10$^7$ M$_{\odot}$ and the other 10$^6$ M$_{\odot}$. It is possible the older cluster started out at a higher mass and lost a large fraction of its mass due to dispersion into the field. How does this compare to sites of massive star formation we can study in detail?

No young 10$^6$ M$_{\odot}$ clusters exist in local group galaxies. However massive sites of star formation (up to roughly 10$^5$ M$_{\odot}$) are accessible in both the Milky Way and the LMC. \citet{gre99} present detailed observations of Hodge 301, a cluster in the 30 Doradus star forming complex. Hodge 301 is between 20-25 Myr old and lies at a distance of 3' ($\approx$ 44 pc) from the young center of 30 Doradus, R136. It is estimated to have been relatively massive with an initial mass of up to 6000 M$_{\odot}$ though less massive than R136 ($\approx$ 30000 M$_{\odot}$).
In the Galaxy, NGC 3603 and the Arches cluster, two of the most massive star forming regions in the Milky Way do not appear to have similar older star formation complexes in their proximity, though both have isolated red supergiants in their environments. \citet{mel05} found the typical separation of young super-star clusters in M82 to be $\approx$ 12 pc though the equivalent age spread of the clusters is not given. \citet{chr08} found that half of their Antennae NIRSPEC cluster sample with 1" resolution contained more than
one star cluster which would indicate a larger typical star cluster separation than in M82. \citet{chr08} also found some sources in their Antennae sample with no obvious
cluster superposition which contained emission lines as well as CO bandheads simultaneously. This raises the question whether some SSC have non-instantaneous bursts
of star formation over the size of the cluster and about the magnitude of this age spread.

Cluster 89/90 illustrates the need to study massive star forming regions in detail at a distance where the individual cluster complexes can be resolved from each other. \citet{neff00} found radii of $\approx$ 3 pc for the thermal radio sources in NGC 4038/9. This distance corresponds to 0.03" in the Antennae. If we assume a typical star cluster
separation of 12 pc, the distance at which a 1" spectrum can resolve this distance is $\approx$ 2.5 Mpc. More distant galaxies make ideal targets for observations with adaptive optics (AO) equipped with integral field unit (IFU) spectrographs.

\section{Summary and Conclusions}

We have modeled the integrated properties of a massive young cluster in the Antennae through its H- and K-band spectrum. Our models use a combination of Starburst99, PMS tracks and field dwarf spectral
standards. We find that the integrated spectrum likely represents two separate star clusters with an age spread of 6-18 Myr with a mass ratio of M$_{old}$/M$_{young}$$\leq$ 0.2 and a total mass of $\approx$ $10^7$ $M_{\odot}$. The IMF in the young cluster is best fit by a \citet{sal55} slope down to the hydrogen burning limit though the cluster IMF is formally consistent with a \citet{cha03} IMF. We find no evidence of a low-mass cutoff in the cluster. Thus we have for the first time directly detected low-mass PMS stars in a young extragalactic super-star cluster.\\

\acknowledgements

The data presented here were obtained with the W.M. Keck Observatory, which is operated as a scientific partnership among the California Institute of Technology, the University of California and the National Aeronautics and Space Administration. The authors wish to recognize and acknowledge the very significant cultural role and reverence that the Mauna Kea summit has always had within the indigenous Hawaiian community. We are most fortunate to have the opportunity to conduct observations from this mountain.
This publication makes use of data products from the Two Micron All Sky Survey, which is a joint project of the University of Massachusetts and the Infrared Processing and Analysis Center/California Institute of Technology, funded by the National Aeronautics and Space Administration and the National Science Foundation.
MRM gratefully acknowledges the support of a Cottrell scholar award from the Research Corporation.

\begin{figure}
\begin{center}$
\begin{array}{cc}
\includegraphics[scale=0.48]{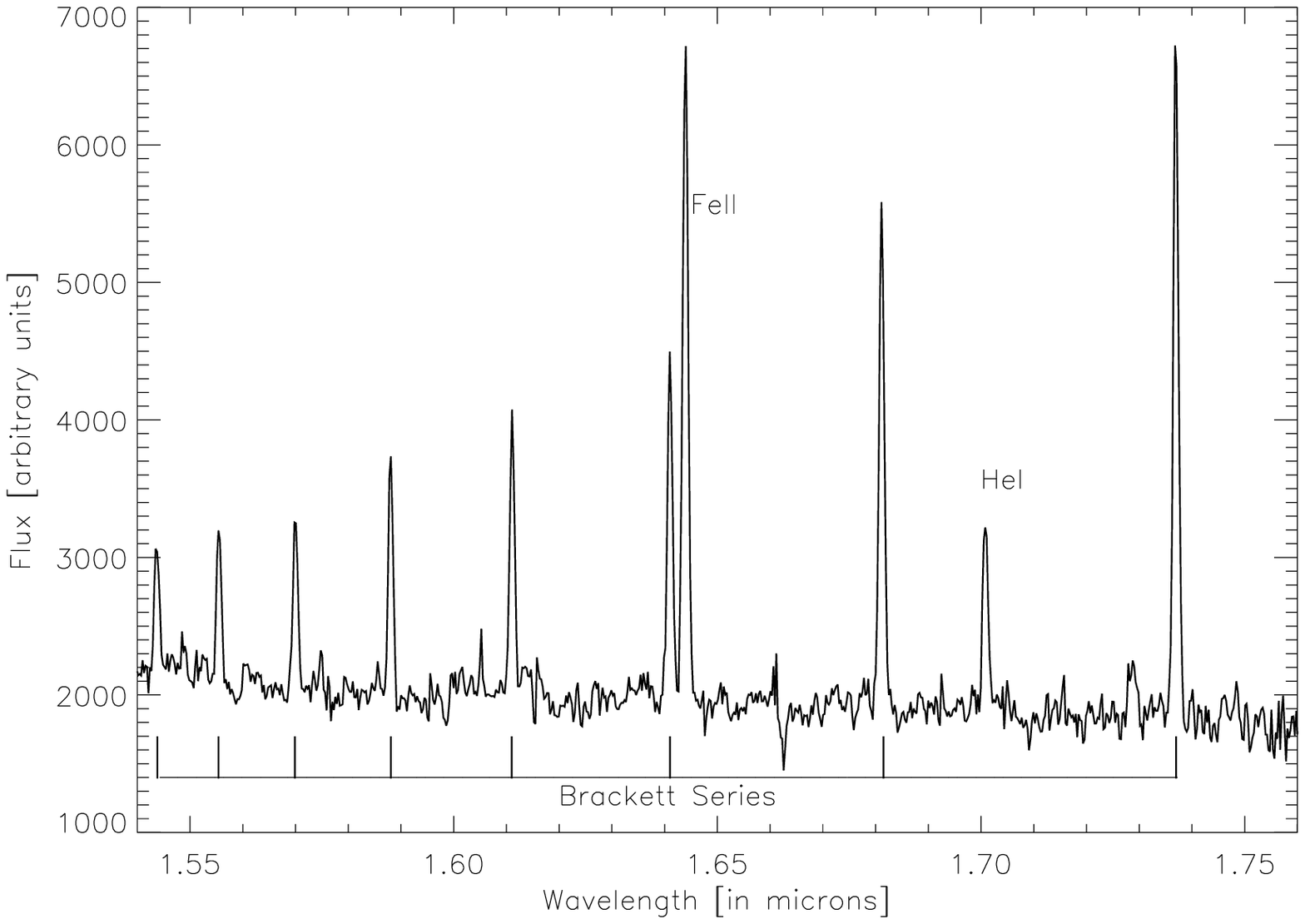} &
\includegraphics[scale=0.48]{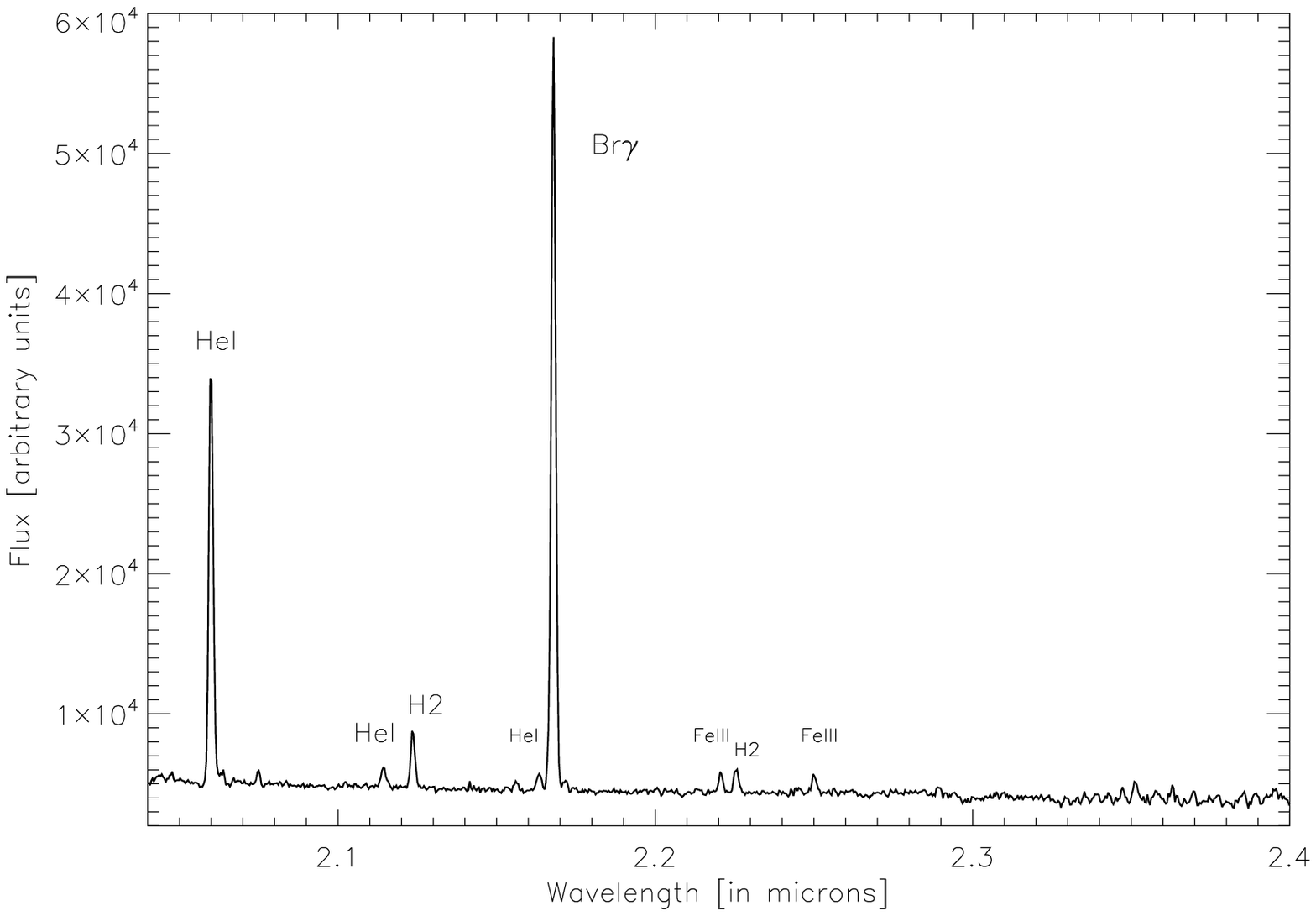}
\end{array}$
\end{center}
\caption{NIRSPEC H- and K-band spectrum of Cluster 89/90. Emission lines which are
indicative of the youth of the cluster are marked. The data have been shifted to restframe wavelengths.}
\end{figure}

\begin{figure}
\begin{center}
\includegraphics[angle=0]{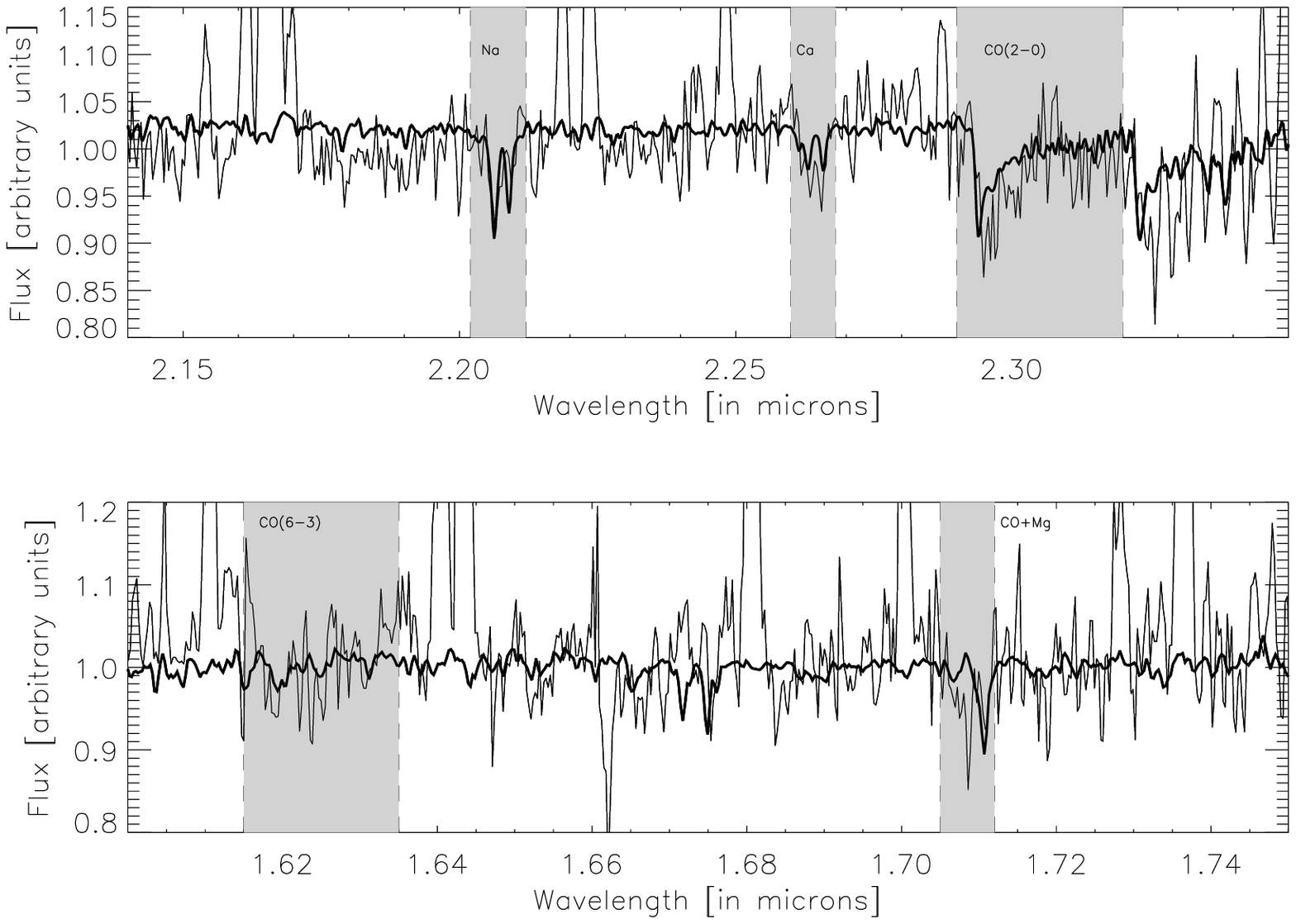}
\end{center}
\caption{Best-fit single age model overlaid over a normalized scaled spectrum of Cluster 89/90. The bands used in our analysis are marked.
The emission lines as well as large parts of the continuum were not included in the modeling. The best-fit reduced $\chi^2$ = 1.40. }
\end{figure}

\begin{figure}
\begin{center}
\includegraphics[scale=0.9]{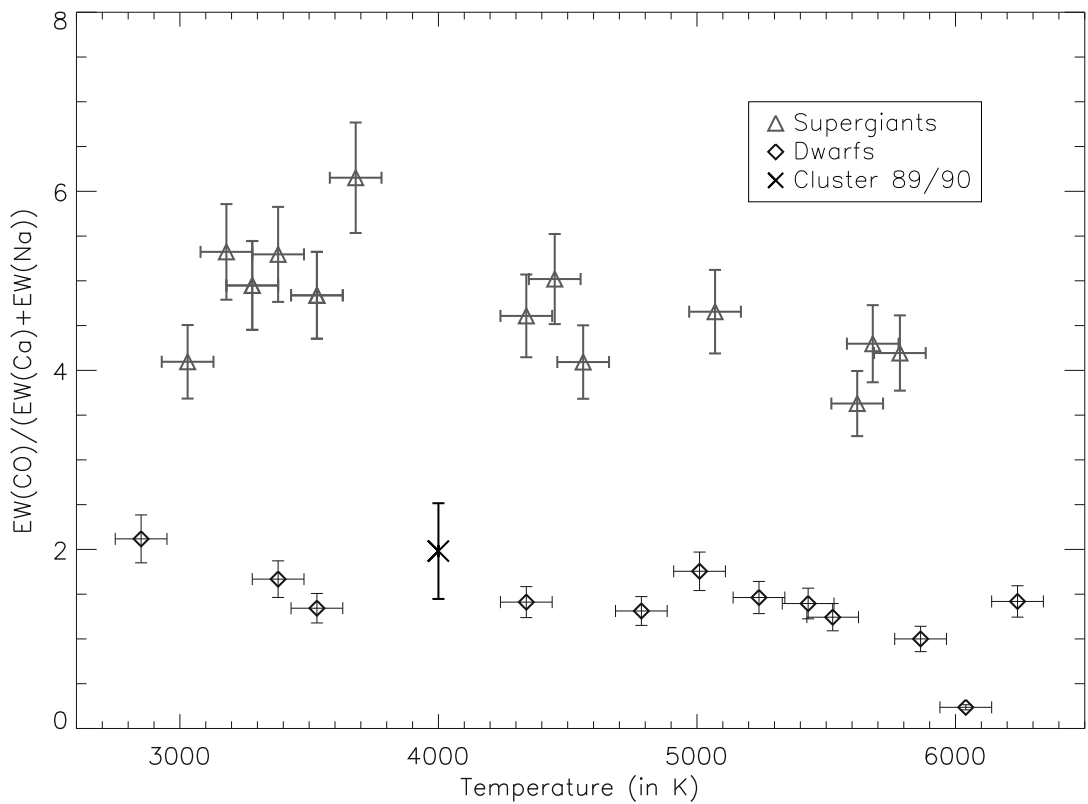}
\end{center}
\caption{Plot of the ratio of equivalent widths of CO(3-1) 2.29 $\micron$, Ca 2.26 $\micron$ and Na 2.20 $\micron$ absorption features
in supergiant and dwarf standards. Overplotted is Cluster 89/90 at an arbitrary temperature. Errors were estimated using
standard error propagation. This shows the clear separation between supergiants
and dwarfs in this index. Cluster 89/90 lies much closer to the dwarf locus rather than the supergiant locus.
This indicates that a supergiant population cannot be solely responsible for the absorption lines present in the spectrum.}
\end{figure}

\begin{figure}
\begin{center}
\includegraphics[angle=0]{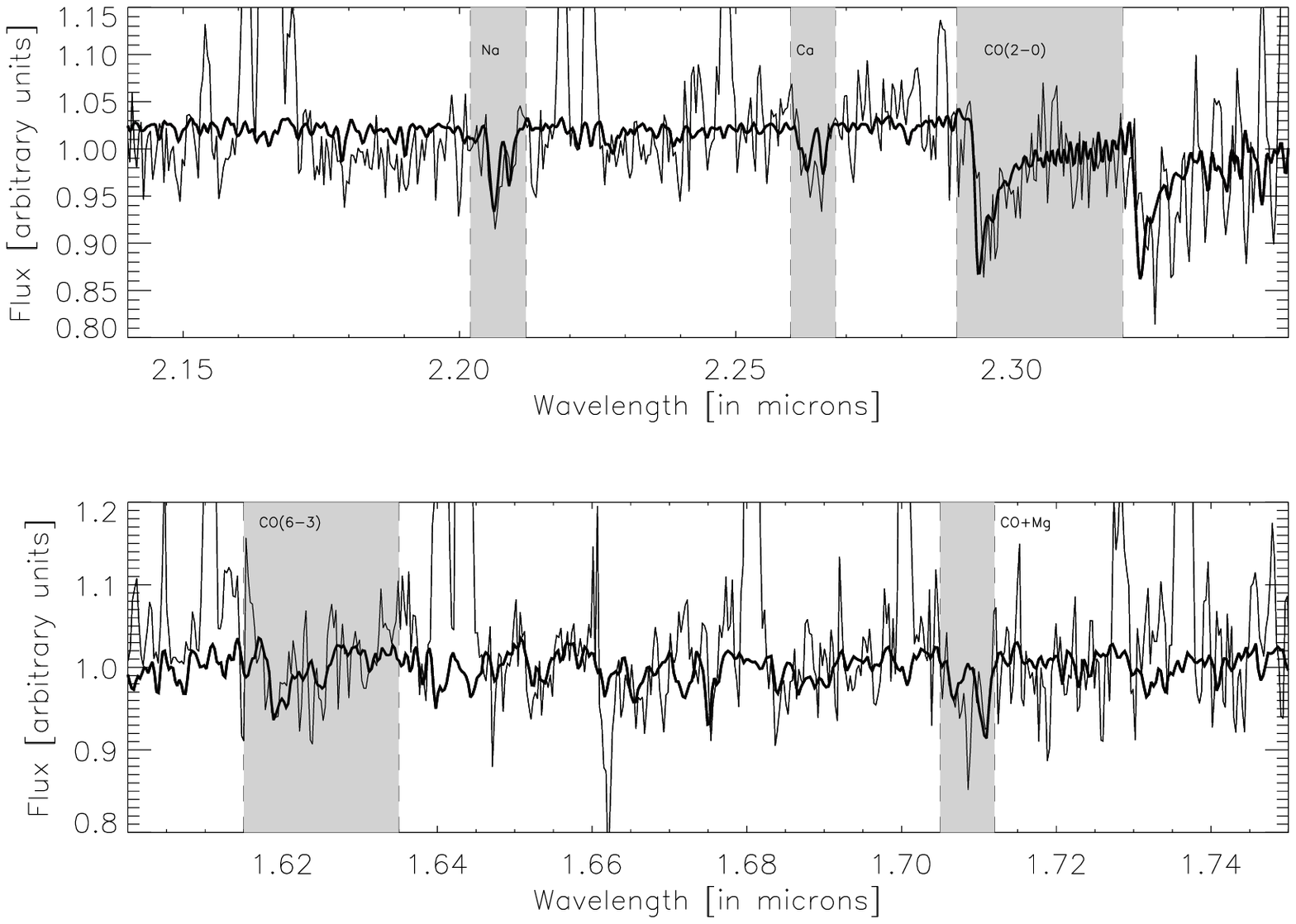}
\end{center}
\caption{Best-fit two-burst model overlaid over a normalized scaled spectrum of Cluster 89/90. The bands used in our analysis are marked. The emission lines as well as large
parts of the continuum were not included in the modeling. The best-fit reduced $\chi^2$ = 1.23. }
\end{figure}

\begin{figure}
\begin{center}$
\begin{array}{ccc}
\includegraphics[scale=0.45]{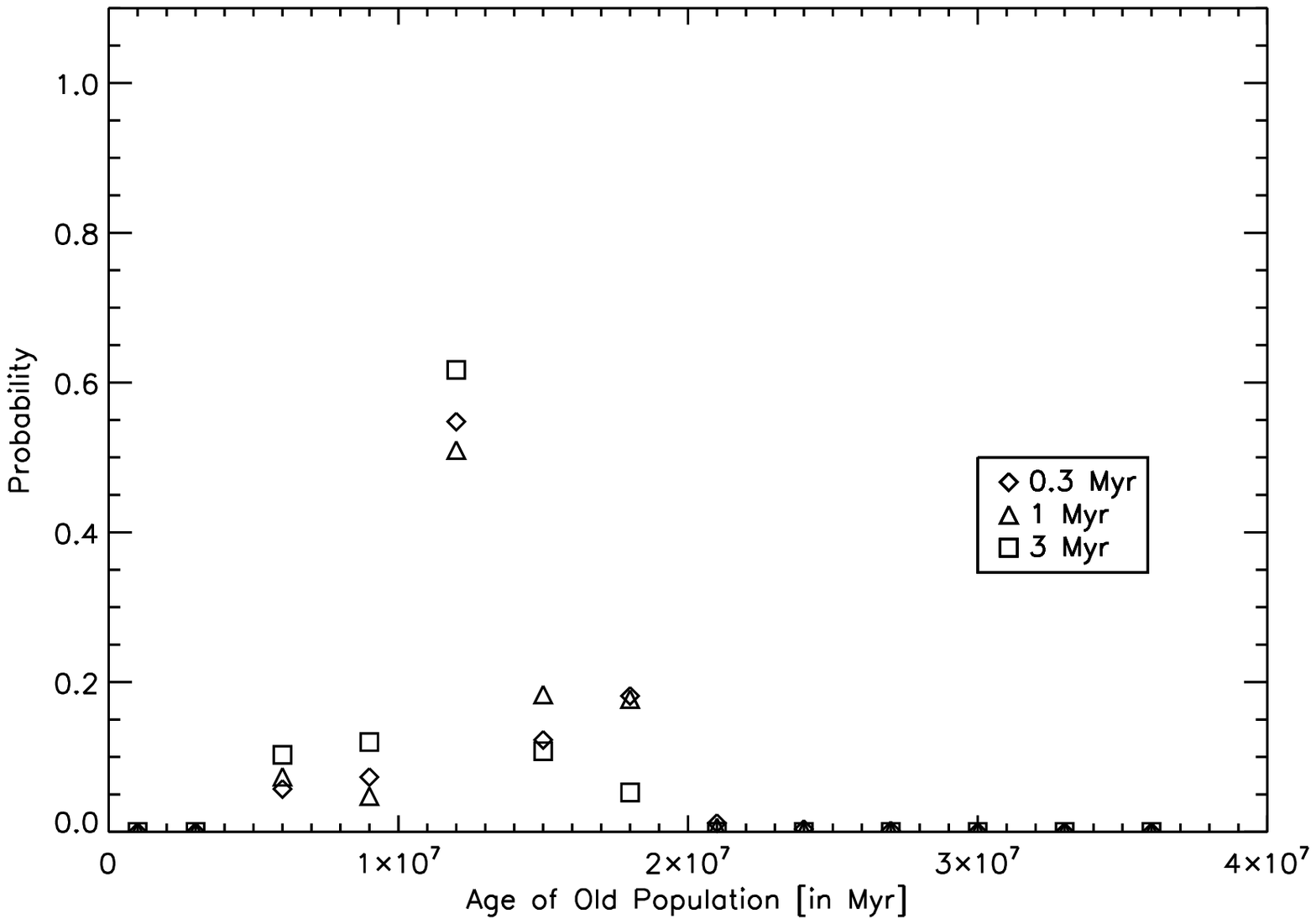} \\
\includegraphics[scale=0.45]{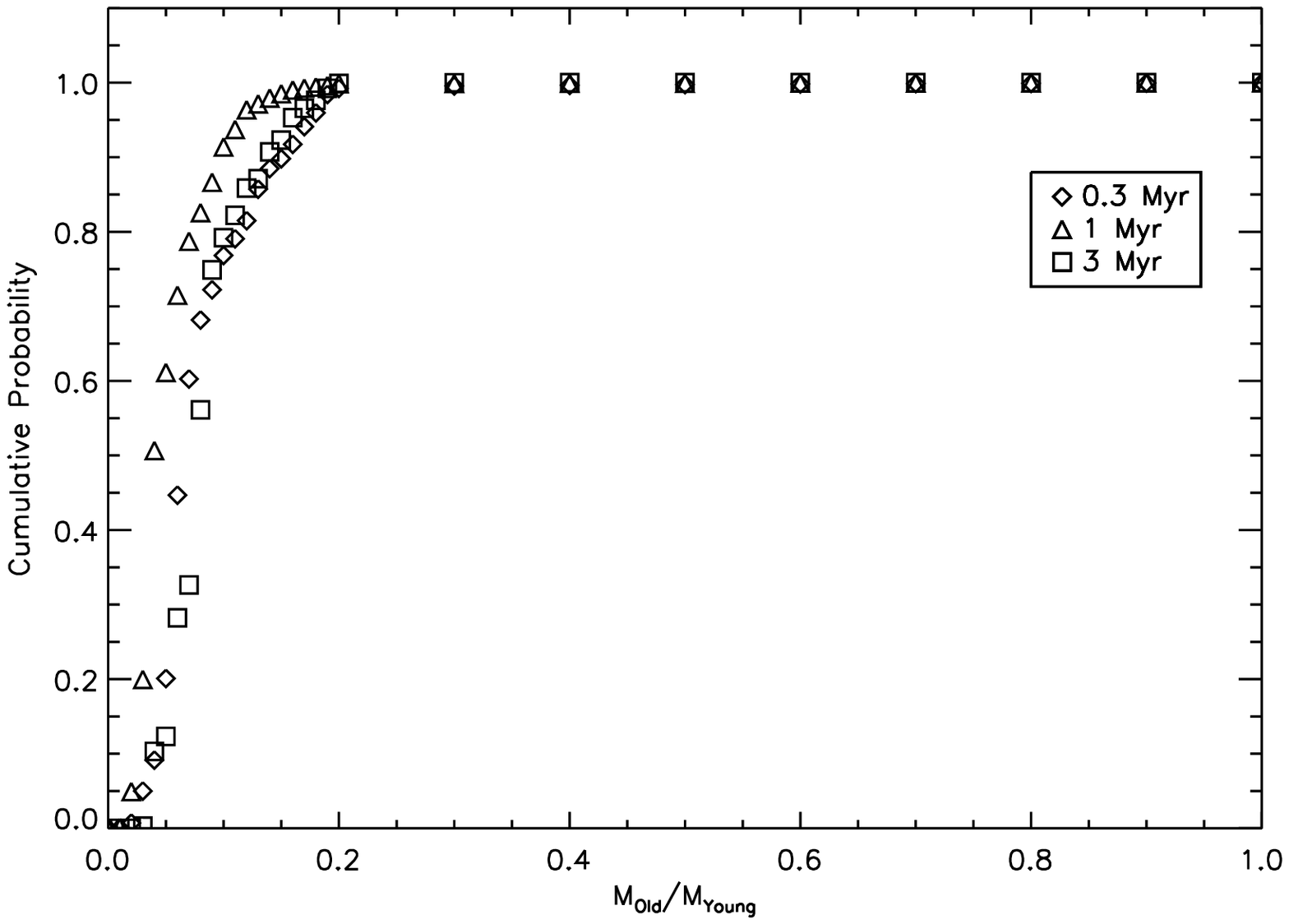} \\
\includegraphics[scale=0.45]{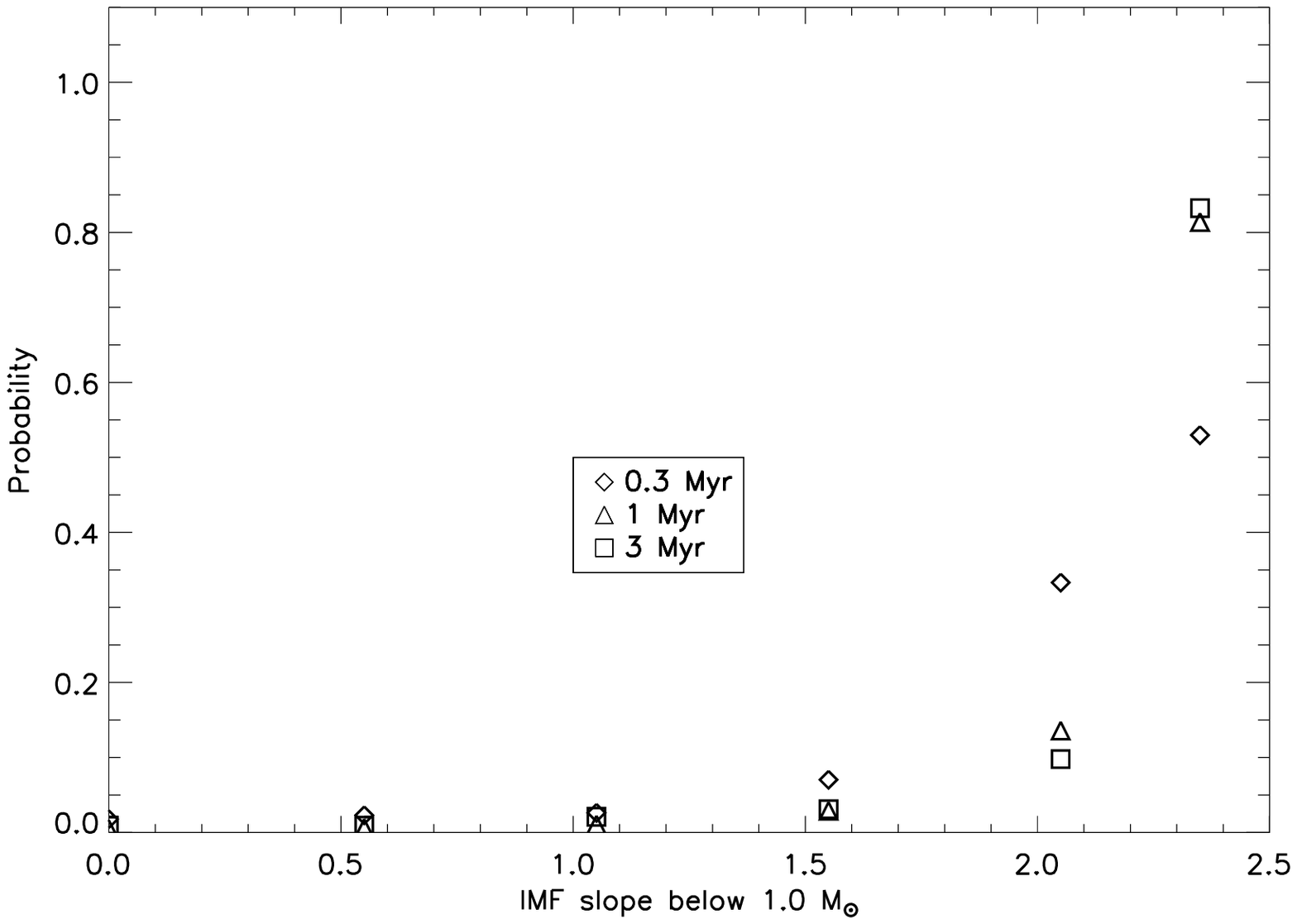} \\
\end{array} $
\end{center}
\caption{Plot of probabilities vs. age (top), mass ratio (middle) and IMF (bottom) for our models. Each plot includes separate data points for all three young burst ages to make certain that our results are not age dependent. Each plot point in a panel represents a summation over all the values of the other two variables and for each panel the probability has been normalized. For clarity the mass ratio is shown as the cumulative probability. Both the IMF and the age plots show well constrained values at the resolution of the models. The mass ratio is less tightly constrained. }
\end{figure}

\clearpage

\begin{deluxetable}{ccccccccc}
\tablecaption{Archival data of cluster 89/90}
\tabletypesize{\scriptsize}
\tablewidth{0pt}
\tablehead{
\colhead{R.A. (J2000.0)} &
\colhead{Dec. (J2000.0)} &
\colhead{F$_{4cm}$ ($\micron$Jy)}\tablenotemark{a}&
\colhead{F$_{6cm}$ ($\micron$Jy)} &
\colhead{m$_V$} \tablenotemark{b}&
\colhead{m$_I$} &
\colhead{m$_J$} \tablenotemark{c}&
\colhead{m$_H$} \tablenotemark{d}&
\colhead{m$_K$} }
\startdata
12:01:54.58 & -18:53:03.42 & 1957 & 2316 & 19.07 & 18.40 & 15.05 & 14.71 & 14.27 \\
\enddata
\tablenotetext{a} {From \citet{neff00}}
\tablenotetext{b} {From Whitmore \& Zhang (2002)}
\tablenotetext{c} {From \citet{bran05} in the 2MASS system}
\tablenotetext{d} {From 2MASS}
\end{deluxetable}

\clearpage

\begin{deluxetable}{cccccc}
\tablecaption{Best-fit Models}
\tabletypesize{\scriptsize}
\tablehead{
\colhead{Model prescription} &
\colhead{Age(in Myr)} &
\colhead{$\alpha$}\tablenotemark{a}&
\colhead{M$_{old}$/M$_{young}$} &
\colhead{M$_{tot}(in M_{\odot})$}\tablenotemark{b}&
\colhead{Reduced $\chi^2$} }
\startdata
Single burst & 1 & -3.0 & .. & 2 x $10^7$ & 1.40 \\
Two bursts (Population A) & 1 & -2.35 & 0.04 & 1.4 x $10^7$ & 1.23 \\
Two bursts (Population B) & 12 & -2.35 & 0.04 & 1.4 x $10^7$ & 1.23 \\
\enddata
\tablenotetext{a} {Power-law slope of the IMF (dN/dm $\propto$ m$^{-\alpha}$)}
\tablenotetext{b} {Flux due to best-fit model as compared with the total K-band flux of cluster 89/90.}
\end{deluxetable}

\end{document}